# Creating and Deleting a Single Dipolar Skyrmion by Surface Spin Twists

Jin Tang[1,2]*, Jialiang Jiang[1], Yaodong Wu[2,3]*, Lingyao Kong[1], Yihao Wang[3], Junbo Li[3], Y. Soh[4], Yimin Xiong[1], Shouguo Wang[5], Mingliang Tian[1,3], and Haifeng Du[3]*

[1]School of Physics and Optoelectronic Engineering, Anhui University, Hefei, 230601, China

[2]School of Physics and Materials Engineering, Hefei Normal University, Hefei, 230601, China

[3]Anhui Province Key Laboratory of Low-Energy Quantum Materials and Devices, High Magnetic Field Laboratory, HFIPS, Chinese Academy of Sciences, Hefei 230031, China

[4]Paul Scherrer Institute, 5232, Villigen, Switzerland

[5]Anhui Key Laboratory of Magnetic Functional Materials and Devices, School of Materials Science and Engineering, Anhui University, Hefei 230601, China

*E-mail: jintang@ahu.edu.cn; wuyaodong@hfnu.edu.cn; duhf@hmfl.ac.cn

**Abstract**

We report deterministic operations on single dipolar skyrmions confined in nanostructured cuboids using in-plane currents. We achieve highly reversible writing and deleting of skyrmions in the simple cuboid without any artificial defects or pinning sites. The current-induced creation of skyrmions is well-understood through the spin-transfer torque acting on surface spin twists of the spontaneous 3D ferromagnetic state, caused by the magnetic dipole-dipole interaction of the uniaxial $Fe_3Sn_2$ magnet with a low-quality factor. Current-induced deletions of skyrmions result from the combined effects of magnetic hysteresis and Joule thermal heating. Our results are replicated consistently through 3D micromagnetic simulations. Our approach offers a viable method for achieving reliable single-bit operations in skyrmionic devices for applications such as random-access memories.

Random access memory (RAM) is ideal for cache memory owing to its high density, high endurance, and non-volatility.[1-4] RAM stores individual data bits in a two-dimensional matrix of cells, which can be accessed randomly. The information carriers in RAM show a great variety and include ferroelectricity, phase change, resistance, and magnetization.[4] In the magnetic RAM (MRAM), uniform opposite magnetizations represent data bits "1" and "0". The data operations in magnetic RAM are achieved by switching the magnetization using field or spin-transfer torque (STT).[3]

Magnetic skyrmions are swirl-like spin textures and possess particular particle-like properties, such as lattice and liquid phases, tunable nanoscale size, and coherent motion driven by small current density.[5-11] These properties enable skyrmions to be high-density bits and can be electrically manipulated with low energy consumption. Magnetic skyrmions are therefore promising information carriers used in building high-performance spintronic devices. In 2013, Fert et al. proposed the skyrmion-based racetrack memory due to the high mobility of skyrmions in nanostripes.[12, 13] Since then, skyrmionic devices have developed many user scenarios with entirely different device geometries including logic computing,[14] probabilistic computing,[15] neuromorphic computing,[16, 17] and RAM[18]. Among them, skyrmion-based RAM has been proposed to offer higher speed and lower energy consumption than traditional magnetic RAM.[18] Toward the functional realization of skyrmion-based RAM, it is required to integrate the full electrical operations on a single skyrmion that is confined in a nanostructured element. Many studies have demonstrated static stabilizations of skyrmions in confined nanostructures, such as disks [19-21] and stripes [22-24], but electrical manipulations of skyrmion numbers in these geometries are still less explored. Electrical methods mainly include voltage applied to insulators and current applied to metals. Recent research indicates that voltage-controlled strain may be an approach to controlling skyrmion numbers in confined disks.[25] However, the deterministic creations and deletions of a single skyrmion in confined nanostructured elements using current are rarely shown.

Here, we demonstrate the deterministic current-induced manipulations of a single dipolar skyrmion in a confined elemental $Fe_3Sn_2$ cuboid, which are understood by STT-driven dynamics of surface spin twists of 3D FM states and magnetic hysteresis. $Fe_3Sn_2$ is a centrosymmetric uniaxial magnet with stacked kagome bilayers and a magnetization easy axis along the [001] axis at room temperature.[26-36] Because of the weak quality factor $Fe_3Sn_2$, the magnetic dipole-dipole interaction leads to surface spin twists of the 3D FM state, which play a role in extrinsic materials defects,[13, 37, 38] in achieving current-induced skyrmion-FM transformations with a zero error rate. Our results provide an approach to controlling skyrmion dynamics in confined nanostructures that could promote skyrmionic device applications.

The magnetic field-driven evolution of magnetic states in a geometrically confined $Fe_3Sn_2$ nanostructure is explored using Lorentz transmission electron microscopy (TEM) (Figure S1).[11] At a zero magnetic field, the magnetic state is a U-shaped stripe domain, which, in turn, transforms to a skyrmion at a magnetic field of $B \sim 329$ mT and a FM state at an increased $B \sim 558$ mT. In contrast, the critical field required for the FM-to-skyrmion transition in the field-decreasing process is $B \sim 470$ mT. Thus, the magnetization curves show hysteresis behavior, *i.e.* skyrmion and FM can coexist in the field range of 470-558 mT. Such a hysteresis is numerically well reproduced (Figure S2) and is the prerequisite for the creation and deletion of skyrmions at a fixed field using current.

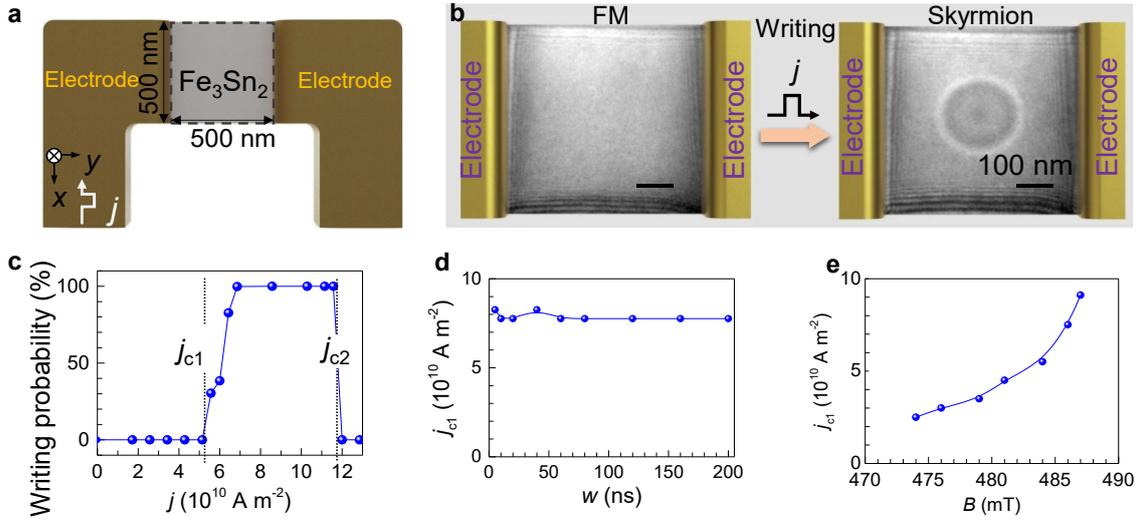

**Figure 1.** Current-induced creation of a single skyrmion. (a) Schematic of $Fe_3Sn_2$ microdevice with the flat surface parallel to the kagome planes. (b) Creation of a skyrmion in $Fe_3Sn_2$ nanostructure using a single 80-ns pulsed current with a current density of $j \sim 8.6 \times 10^{10}$ A m$^{-2}$. $B \sim 484$ mT. The defocused value used for the Lorentz TEM imaging is 1000 μm. (c) Current density $j$ dependence of writing probability at $B \sim 484$ mT. (d) Pulse duration dependence of threshold current density $j_{c1}$ at $B \sim 484$ mT. (e) Magnetic field dependence of $j_{c1}$.

Setting FM as the initial state in the hysteresis range at $B = 484$ mT, we study the current-induced dynamics. A single 80-ns pulsed current is applied to the microdevice (Figure 1a). FM remains as the current density of $j$ is below a threshold value of $j_{c1} \sim$ $5.6 \times 10^{10}$ A m$^{-2}$, above which a skyrmion is created (Figure 1b and 1c). The writing skyrmion probability increases with the increase of current density. In the current density range of $j \sim (6.6\text{-}11.6) \times 10^{10}$ A m$^{-2}$, the success rate of writing skyrmions using a single pulsed current reaches 100%. However, when the pulsed current density is above a threshold value of $j_{c2} \sim 12.0 \times 10^{10}$ A m$^{-2}$, the skyrmion cannot be created, and the writing probability decreases to 0 suddenly.

The current-induced effects mainly include non-thermal STT, Joule thermal heating, and Oersted field effects. For the Joule thermal heating effect, the thermal energy $E_{\text{therm}}$ induced by the current is expressed by $j^2w$, where $w$ is the pulse duration. Thus, the thermal-induced creation of skyrmions must be dependent on pulse duration $w$. However, the threshold density $j_{c1}$ of current required for creating a skyrmion is not dependent on $w$ (Figure 1c), which excludes the Joule thermal heating as the physical origin of creating a skyrmion. The shortest current pulse we have achieved for skyrmion writing is 5 ns. Further, the current-induced Oersted field is dependent on current orientation. However, our results demonstrate that the current-induced skyrmion creation is also realized by altering the orientation of the current or the magnetic field (Figure S3b), which excludes the Oersted field effect. The above analysis suggests that STT is responsible for the current-induced creation of single skyrmions in the confined cuboid. The threshold current density $j_{c1}$ is very sensitive to magnetic field $B$. As shown in Figure 1E, the threshold density $j_{c1}$ increases from $\sim 2.5 \times 10^{10}$ A m$^{-2}$ at $B = 474$ mT to $\sim 9.1 \times 10^{10}$ A m$^{-2}$ at $B = 488$ mT.

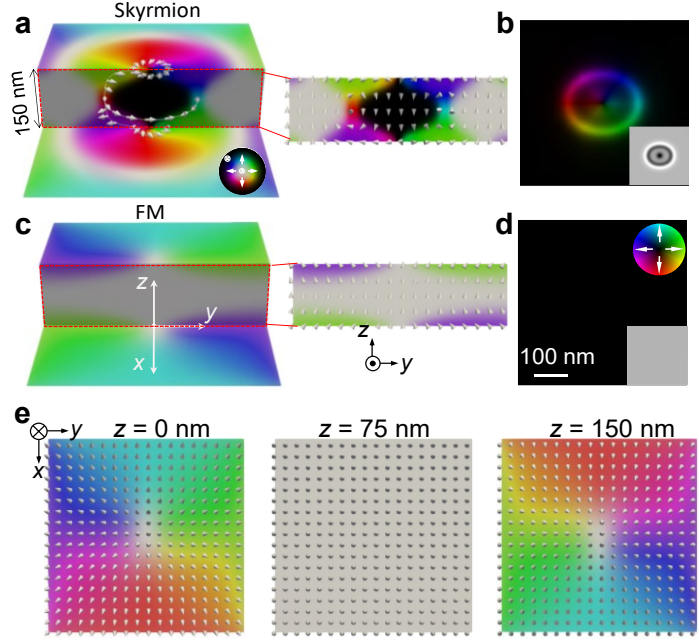

**Figure 2.** Surface twists. (a) Simulated 3D skyrmion state in a 150-nm thick uniaxial magnetic cuboid. (b) Overall in-plane magnetization mapping of the 3D skyrmion. (c) Simulated 3D ferromagnetic state in a 150-nm thick uniaxial magnetic cuboid. (d) Overall in-plane magnetization mapping of the 3D FM. (e) Spin configurations in the top surface, middle layer, and bottom surface of the 3D FM.

The competition among perpendicular magnetic anisotropy, dipole-dipole interaction, and exchange interactions is the origin of stripes and dipolar skyrmions (also called skyrmion bubbles or type-I bubbles) in $Fe_3Sn_2$.[26-36] Because of the absence of chiral interaction, we can simultaneously stabilize dipolar skyrmions with clockwise and counterclockwise rotations in the $Fe_3Sn_2$ nanostructure (Figure S4). The quality factor $\eta$ is an important index to describe the stability of magnetic skyrmion bubbles and is defined as $\eta = \frac{2K_u}{\mu_0 M_s^2}$, where $K_u$ is the perpendicular anisotropy, $M_s$ is the saturated magnetization, and $\mu_0$ is the vacuum permeability. $Fe_3Sn_2$ has a weak quality factor ($\eta \approx 0.22$) at room temperature. Thus, the magnetic configurations in $Fe_3Sn_2$ reveal obvious depth-modulated spin twists because of the demagnetization field effect.[26] The skyrmion therein shows a 3D structure with Néel-type spin twists at the surface and Bloch-type twisting in the interior (Figure 2a), as is also experimentally identified in previous work from the average in-plane magnetizations (Figure 2b).[26] It is worth noting that Néel-type spin twists near the surface do not arise in ultrathin ferromagnetic films, even when a weak quality factor is present. Instead, a relatively weak quality factor facilitates the stabilization of dipolar topological magnetic configurations with arbitrary topological charges $Q$.[39]

In chiral magnets, skyrmions annihilate at magnetic fields but non-uniform magnetizations known as conical magnetizations can remain.[6, 40-42] In uniaxial magnets without DM interactions, magnetic states for the absence of skyrmions at magnetic fields are not well shown. Here, setting a uniform FM state in the centrosymmetric $Fe_3Sn_2$ magnets without DM interactions as the initial state in simulations, a relaxed stable state is obtained. We find that the uniform FM state cannot be stabilized at a low field region and transforms to a special 3D FM state, as shown in Figure 2c. The magnetization in the interior layers of the 3D FM state remains uniform but magnetizations in near-surface layers turn to vortex-like spin textures. The depth-modulated spin twists of the 3D FM state are also attributed to the magnetic dipole-dipole interaction. Our simulations reveal that surface twists near the center of the 3D FM gradually diminish and transition into a nearly uniform FM when the magnetic field exceeds ~500 mT (Figure S5).

Despite the non-uniform surface domains of 3D FM, the average in-plane magnetization of 3D FM altogether cancels out. Thus, no Fresnel contrast is observed for the 3D FM (Figure 2D) which is consistent with our experiments (Figure 1b). The existence of such a 3D FM state is also experimentally identified in a previous study from anisotropic magnetoresistance measurements,[43] which reveals a resistance change in a field region where the FM state is kept. Such a resistance change is attributed to the non-uniform surface spin twists and agreed with numerical simulations.[43] Because the spatial magnetization gradients $\nabla \mathbf{m}$ of the surface domains in 3D FM are nonzero, the 3D FM could have dynamical responses to a current-induced STT torque even without the consideration of material defects.

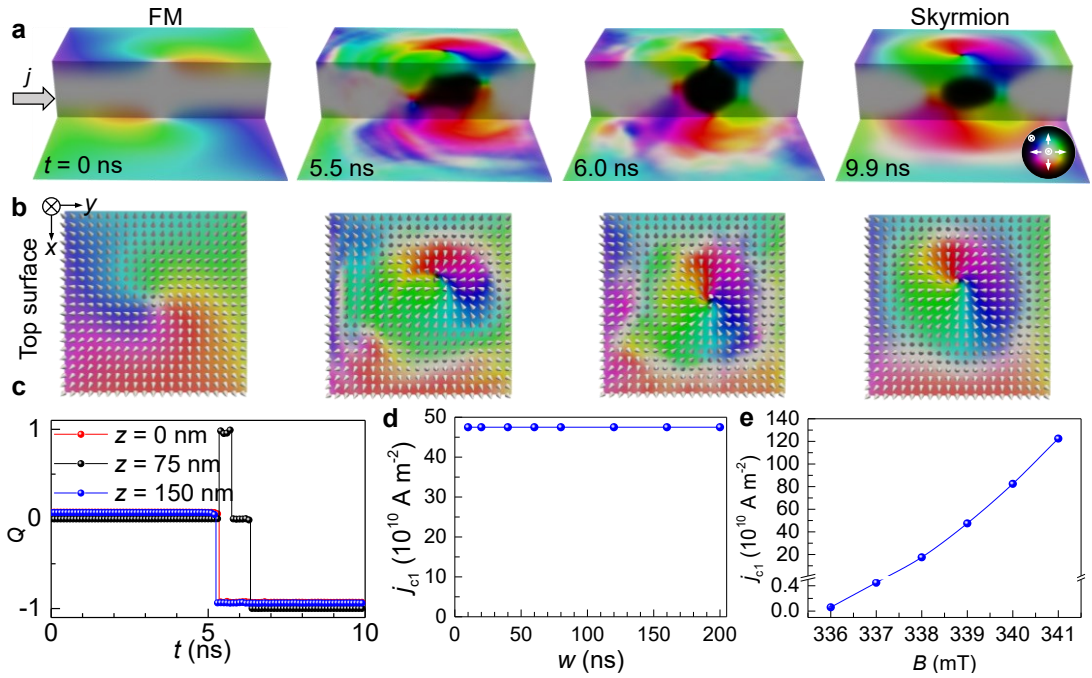

**Figure 3.** STT-induced creation of a skyrmion. (a) Simulated snapshots of overall 3D configurations, (b) 2D magnetic configurations at the top surface, during the course of the STT-driven creation of a single skyrmion from FM. Simulated current density $j = 50 \times 10^{10}$ A m$^{-2}$. (c) Time $t$ dependence of $Q$ during the FM-to-skyrmion transformation in the layers at the depth $z = 0$, 75, and 150 nm. (d) Simulated pulse

duration $w$ dependence of $j_{c1}$. (e) Simulated pulse duration $w$ dependence of $j_{c2}$.

The microscopic physical mechanisms for the current-induced creation of single skyrmions are explored using micromagnetic simulations based on measured magnetic parameters of Fe$_3$Sn$_2$,[26, 27, 44] where skyrmion creation can be well explained by the STT effect, as shown in Figure 3 and Supplemental Video 1. In our simulation, at the magnetic field of $B = 339$ mT which is far smaller than the threshold saturated magnetization field $B = 470$ mT, FM is the metastable phase with higher energy than that of skyrmion (Figure S2). After applying a current with a density of $50 \times 10^{10}$ A m$^{-2}$, the vortex-like surface spin twists of 3D FM are distorted, and non-uniform magnetizations appear in the middle layers at time $t = 5.0$ ns. At $t = 5.2$ ns, a seed domain with magnetization antiparallel to the field is nucleated to decrease the dipole-dipole interaction. Such a seed domain quickly expands its size and expands over to all layers to form a distorted magnetic bubble at $t = 5.5$ ns. The Bloch lines in the middle bubble domains, which are not supported at a perpendicular field, may appear and contribute transient $Q = 1$ and $Q = 0$ bubbles (Figure 3c). However, these non-equilibrium states with high-energy Bloch lines are quickly annihilated when continuously applying the current, yielding the lowest energy topologically non-trivial skyrmion at $t = 6.5$ ns. In the simulation, the threshold current density $j_{c1} = 47.5 \times 10^{10}$ A m$^{-2}$ is higher than the experimental value because the temperature effect is not considered in simulations. Because the nucleation of skyrmions happens in about 5 ns and the skyrmion with the lowest energy remains even if the current is still on for $w >$ 5 ns, the STT-induced creation of skyrmions is determined only by the pulse amplitude for the pulse duration $w > 5$ ns. In our simulation, the threshold current density $j_{c1}$ is independent of $w$ (Figure 3d) which is consistent with our experiments (Figure 1d), proving that skyrmion creation is unambiguously associated with STT effects. In addition, since the skyrmion is more stable than FM in the low magnetic field region, the threshold density $j_{c1}$ is very sensitive to the magnetic field (Figure 3e) and agrees with the experiments (Figure 1e). Our experimental results are all well reproduced in our simulations, suggesting the reliable physical origin of STT-induced dynamics of surface spin twists as the creation of single skyrmions.

STT cannot lead to FM-to-skyrmion transformation because of the zero spatial magnetization gradient ($\nabla \mathbf{m} = 0$) for uniform FM,[45, 46] Thus, in previous studies of writing skyrmions from FM, non-uniform magnetizations induced by artificial defects or localized impurities must be built as seed domains to create skyrmions by STT effects.[13, 38, 45, 47-50] Our results suggest that non-uniform surface spin twists can naturally form because of the intrinsic magnetic dipolar-dipolar interactions in uniaxial magnets without specified material disorders. In low-field regions, the metastable high-energy 3D FM with surface spin twists can thus be taken as the seed domain to form the low-energy skyrmion in the stimuli of STT.

It could be pointed out that magnetic dipole-dipole interactions are also important in stabilizing skyrmions in multi-layer films,[51, 52] chiral magnets with $D_{2d}$ symmetries,[53, 54] and 2D magnets[55, 56]. Similar depth-modulated Bloch-Neel spin twists contributed by magnetic dipole-dipole interactions have also been shown,[51, 52] which predicts the stabilization of such 3D FM states in a wide variety of materials.

Thus, STT-driven dynamics of surface twists of 3D FM states are potentially applicable for the creation of antiskyrmions in $D_{2d}$ magnets, and skyrmion bubbles in multilayers and 2D magnets.

By setting the skyrmion as the initial state in the hysteresis range at a magnetic field of $B$ = 484 mT, an 80-ns pulsed current with a density above the critical value of $j_{c2} \sim 12.0 \times 10^{10}$ A m$^{-2}$ will delete the skyrmion (Figure 4). The Joule heating effect is considered the mechanism of the skyrmion deletion at a high current pulse. A pulsed current with a small duration $w$ requires a large current density $j$ to induce equivalent Joule heating. Thus, the threshold current density $j_{c2}$ required for heating-induced skyrmion deletion would decrease as $w$ increases, which is experimentally verified from the experimental $w$-$j_{c2}$ relation (Figure S3a).

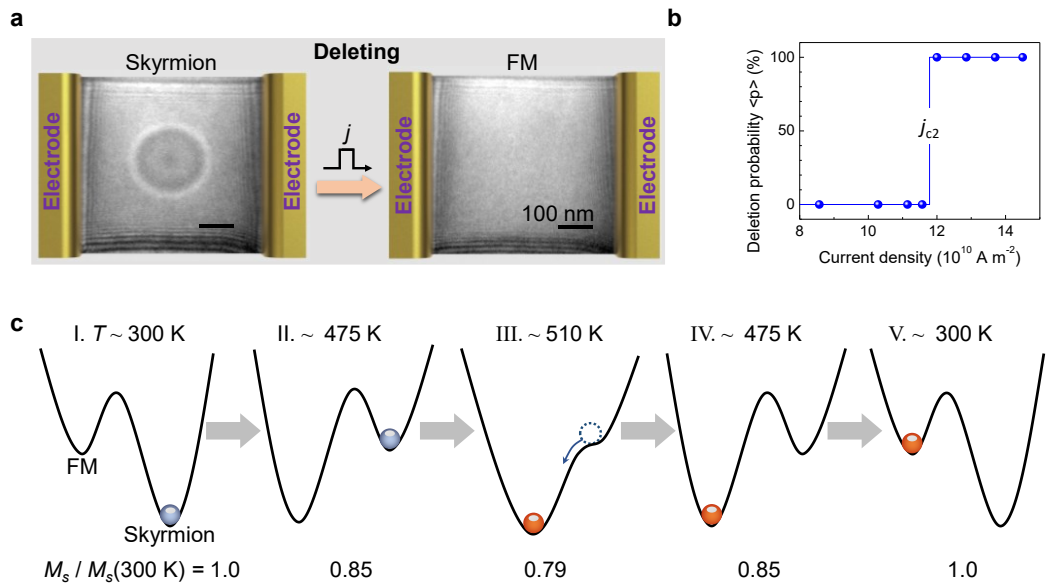

**Figure 4.** Deleting a single skyrmion. (a) Annihilation of a skyrmion in Fe$_3$Sn$_2$ nanostructure using an 80-ns pulsed current with density $j \sim 12.0 \times 10^{10}$ A m$^{-2}$ at $B \sim$ 484 mT. (b) Current density $j$ dependence of deleting probability. (c) Schematic

energy density profiles under a cycle of varying temperatures of 300, 475, 510, 475, and 300 K show the transition from a skyrmion to FM.

A previous study have demonstrated that a 100-ns current density of 3.4 ×10$^{10}$ A m$^{-2}$ can lead to a rising temperature of ~180 K,[30] which contributes to a reduction of the saturated magnetization $M_s$ by ~20%.[28] Micromagnetic simulated magnetic evolutions driven by varied $M_s$ in the Fe$_3$Sn$_2$ cuboid are explored by setting the magnetic field, exchange interaction, and anisotropy constant. In these simulation cases, the threshold saturated magnetization field decreases when $M_s$ decreases and increases when $M_s$ increases, which satisfies the resultant temperature effect in experiments. The simulated results show that the skyrmion gets smaller with the decreasing saturated magnetization at a fixed magnetic field $B$ = 339 mT (Supplemental Video 2). When the saturated magnetization is reduced by 21%, the skyrmion is unstable and transforms to FM according to our simulations. Once the FM forms, it will persist even if the saturated magnetization is later recovered. The heating-induced deletion of skyrmions is summarized from the schematic energy landscape (Figure 4c), this is a typical routine issue and similar to the magnetization hysteresis loops in the Stoner-Wohlfarth model describing the uniform rotation of single-domain magnetic particles.[57]

Keeping in mind the creation and annihilation of single skyrmions using two discrete pulsed currents with low and high current densities, respectively, we further performed combined operations to test the repeatability of skyrmion-FM transformations (Figure 5 and Supplemental Video 3). A low current density of ~8.6 × $10^{10}$ A m$^{-2}$ and a high current density of ~12.0 × $10^{10}$ A m$^{-2}$ are chosen to create and annihilate a single skyrmion, respectively. We found that such operations can be deterministic (Supplemental Video 3 and Video 4) even after the application of 10 000 cycles.

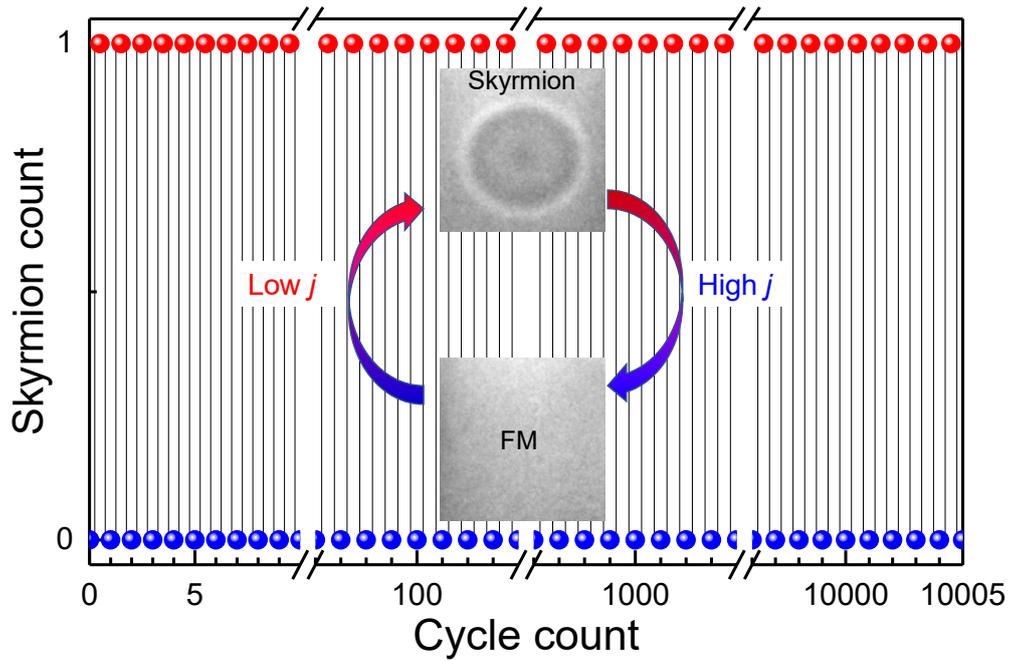

**Figure 5.** Deterministic writing and deleting of single skyrmions. Cycle count dependence of skyrmion creation and deletion at $B$ ~ 484 mT. Each cycle count includes a low current pulse with a density of ~8.6 × $10^{10}$ A m$^{-2}$ and a subsequent high current pulse with a density of ~12.0 × $10^{10}$ A m$^{-2}$. The time interval between two discrete 80-ns pulsed currents is 1.3 s.

Skyrmion-FM transformations can also be achieved in thinner Fe$_3$Sn$_2$ cuboids with a thickness of ~100 nm (Figure S6). However, when the thickness of the Fe$_3$Sn$_2$ cuboids drops below ~70 nm, the in-plane shape anisotropy begins to dominate over the perpendicular magnetocrystalline anisotropy, leading to the instability of dipolar skyrmions and the emergence of soft vortex states (Figure S7).

The skyrmion-FM transformation, an immediate topic of skyrmionic devices, can be applied in writing/deleting operations and demonstrated using hydrogen adsorption/desorption,[58] optical pulses,[59, 60] etc. In spintronic devices, currents could be one of the preferred techniques in achieving writing/deleting operations. Niklas et al. were able to achieve reversible writing-deleting operations on single skyrmions using spin-tunneling currents at extremely low temperatures.[9] Recent current-induced writing/deleting skyrmions at room temperature in one device need to change the external fields before switching to the other operations,[16, 61, 62] have error rates,[37] or are assisted by intrinsic material defects which however are not definiteness[47]. Here, we demonstrate the high repetition in skyrmion-FM transformations in elemental nanostructures relevant to devices. The intricate electronic topology, featuring the spin-gapped nodal loop in the Kagome magnet $Fe_3Sn_2$,[63] leads to an enhancement of the anomalous Hall effect. This could potentially pave the way for effective electrical detection of skyrmions in $Fe_3Sn_2$-based devices. Furthermore, considering the successful electrical detection of a single dipolar skyrmion in a specific nanostructured cell,[43] the proof-of-concept of skyrmion-based MRAM with complete electrical functionalities (including writing, deleting, and reading) has been achieved and shows potential for future topological device applications.

In summary, we have successfully achieved the deterministic creation and deletion of a single skyrmion in an elemental $Fe_3Sn_2$ cuboid using single ns-scale pulsed currents, which are delivered at an extremely high repetition rate. We report a special 3D FM state with surface twists which can be used for STT-driven creation of skyrmions in the simple nanostructure cuboid. The skyrmion-to-FM transformation is induced by magnetization hysteresis and current-induced Joule thermal heating. Our findings provide an approach for creating spatial non-uniform magnetizations for STT-controlled skyrmion dynamics, which can be applied to a wide range of skyrmionic materials with impactful dipolar-dipolar interactions, such as multi-layer films,[51, 52] $D_{2d}$ magnets,[53, 54] and 2D magnets[55, 56]. The controlled skyrmion-FM transformations in confined cuboids using in-plane currents can also serve as a valuable guide in the design of random-access memory applications.

## Methods

**Fabrication of $Fe_3Sn_2$ microdevices**

Thin $Fe_3Sn_2$ microdevices with standard four electrical contacts for resistance measurements were fabricated from the bulk using a standard lift-out method [11, 26, 27], with a focus ion beam and scanning electron microscopy dual beam system (Helios Nanolab 600i, FEI). The length and width of the nanostructure are designed to be both ~500 nm, where one skyrmion bubble can be stabilized at most.

**TEM measurements**

We used in-situ Lorentz-TEM (Helios F200X, FEI) to investigate the current-induced magnetic domain dynamics in the $Fe_3Sn_2$ nanostructure. The current pulses with pulse durations varying from 5-200 ns were provided by a voltage source (AVR-E3-B-PN-AC22, Avtech Electrosystems). All experiments were performed at room temperature.

**Micromagnetic simulations**

The zero-temperature micromagnetic simulations were performed using MuMax3.[44] We consider the Hamiltonian exchange interaction ($A$) energy, uniaxial magnetic anisotropy ($K_u$) energy, Zeeman energy, and dipole-dipole interaction energy.[44] A Zhang-Li spin-transfer torque is considered for simulating current-driven dynamics.[46] Simulated magnetic parameters ($K_u$ = 54.5 kJ m$^{-3}$, $M_s$ = 622.7 kA m$^{-1}$, and $A$ = 8.25 pJ m$^{-1}$) are set based on the measured parameters of Fe$_3$Sn$_2$ at room temperature [26, 27]. The Gilbert damping and non-adiabatic parameters in the dynamic simulations are 0.05 and 0.0, respectively. To understand heating effects induced by the current, the varied simulation parameters need to reflect similar resultant temperature effects, *i.e.*, the saturated magnetization field decreases when temperature increases. We thus set that $M_s$ decreases from 622.7 to 435.89 kA m$^{-1}$ and then $M_s$ recovers back to 622.7 kA m$^{-1}$; the equilibrium magnetic state was obtained using the conjugate-gradient method. We set the length, width, and height of simulated geometry as 500 nm, 500 nm, and 150 nm (or 60 nm), respectively. The cell size was set at 2 × 2 × 3 nm$^3$.

■ **ASSOCIATED CONTENT**

**Data Availability Statement**

The data that support the findings of this study are available from the corresponding authors upon reasonable request.

**Supporting Information**

The Supporting Information is available free of charge on the ACS Publications website at http://pubs.acs.org.

**Figures S1 to S7.** Magnetic phase diagram in increased and decreased field processes.

Simulated field-driven magnetic evolution in the $Fe_3Sn_2$ cuboid. Pulse duration and magnetic field dependence of threshold current densities required for the creation $j_{c1}$ and deletion $j_{c2}$. Dipolar skyrmions with two styles of rotation in the $Fe_3Sn_2$ cuboid. Simulated field-driven magnetic evolution of ferromagnet with surface twists. Magnetic evolution and current-driven dynamics in a 100-nm thick $Fe_3Sn_2$ cuboid. Instability of skyrmion and emergence of soft vortex in ~70 nm-thick $Fe_3Sn_2$ cuboid.

**Video S1 to S4.** Simulated STT-driven FM-to-skyrmion transformation. Simulated skyrmion-to-FM transformation. Writing and deleting single skyrmions using single 80-ns pulsed currents with densities of 8.6 and $12 \times 10^{10}$ A m$^{-2}$, respectively. Writing and deleting single skyrmions using single 80-ns pulsed currents with densities of 10.0 and $12.5 \times 10^{10}$ A m$^{-2}$, respectively.

■ **AUTHOR INFORMATION**

Corresponding Authors

*(J.T.) Email: jintang@ahu.edu.cn; *(Y.W.) Email: wuyaodong@hfnu.edu.cn; *(H.D.) Email: duhf@hmfl.ac.cn

Authors

**Jin Tang-** https://orcid.org/0000-0001-7680-8231

**Author Contributions**

H.D. and J.T. supervised the project and conceived the experiments. Y. Wang, J.L., and Y.X. synthesized the $Fe_3Sn_2$ bulk crystals. J.T. fabricated the $Fe_3Sn_2$ microdevices and performed the TEM measurements with the help of J.J. and Y. Wu. J.T. performed the simulations with the help of L.K.. J.T., Y.W., and H.D. wrote the manuscript with input from all authors. All authors discussed the results and contributed to the manuscript.

■ ACKNOWLEDGMENTS

This work was supported by the National Key R&D Program of China, Grant No. 2022YFA1403603; the National Natural Science Funds for Distinguished Young Scholars, Grant No. 52325105; the National Science Foundation for Excellent Young Scholars, Grant No. 12422403; the Natural Science Foundation of China, Grants No. 12174396, 12104123, and 12241406; the Anhui Provincial Natural Science Foundation, Grant No. 2308085Y32; the Natural Science Project of Colleges and Universities in Anhui Province, Grants No. 2022AH030011 and 2024AH030046; the Strategic Priority Research Program of Chinese Academy of Sciences, Grant No. XDB33030100; CAS Project for Young Scientists in Basic Research, Grant No. YSBR-084; Systematic Fundamental Research Program Leveraging Major Scientific and Technological Infrastructure, Chinese Academy of Sciences, Grant No. JZHKYPT-2021-08; Anhui Province Excellent Young Teacher Training Project Grant No. YQZD2023067; the 2024 Project of GDRCYY (No. 217, Yaodong Wu); and the China Postdoctoral Science Foundation Grant No. 2023M743543.

■ REFERENCES

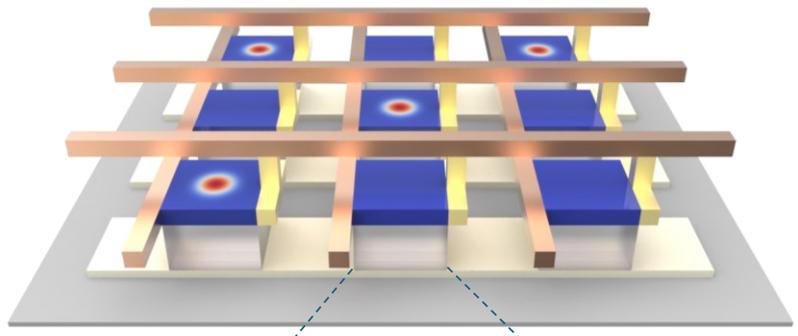
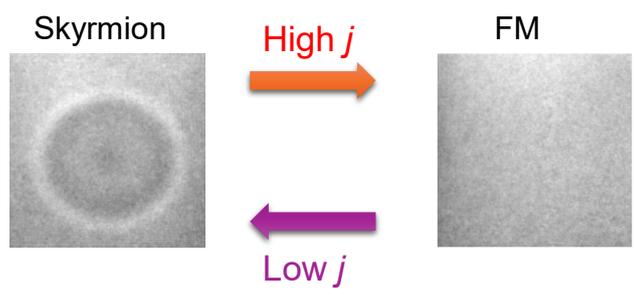

**For Table of Contents Only**